\documentclass[aps,prb,reprint,amssymb,amsmath,superscriptaddress,nobibnotes,citeautoscript]{revtex4-1}

\usepackage{graphicx}
\usepackage{color}

\usepackage{hyperref}

\allowdisplaybreaks

\usepackage{siunitx}
\usepackage{bm}

\newcommand{\beq}[1]{\bm{#1}}  

\newcommand{\Qo}{Q_0}               
\newcommand{\sdo}{\tilde{{\bf\Lambda}}_{\rm sd}} 
\newcommand{\upsh}{\hat{\bm{u}}_{\rm PSH}}

\newcommand{\umlaut}{\"}
\newcommand{\psh}{\beq{\Omega}_{\rm PSH}}
\newcommand{\3}{\beq{\Omega}_{\rm 3}}
\newcommand{\parag}{\\\indent}

\newcommand{\y}{\hat{\beq{y}}}
\newcommand{\bo}{\beq{\Omega}}

\usepackage[normalem]{ulem}

\newcommand{\refchap}[1]{\hypersetup{linkcolor=RoyalRed}\ref{#1}} 
\newcommand{\reffig}[1]{\hypersetup{linkcolor=RoyalRed}\ref{#1}}
\newcommand{\refeq}[1]{\hypersetup{linkcolor=RoyalRed}\eqref{#1}}

\definecolor{olive}{rgb}{0.000 ,0.5216 ,0.1373}
\definecolor{RoyalRed}{RGB}{157,16, 45}
\definecolor{skyblue}{RGB}{0 ,109 ,255}
\definecolor{RoyalGreen}{RGB}{37, 102, 43}
\definecolor{purple}{rgb}{0.7412 ,0.000 ,1.000}
\definecolor{black}{rgb}{0,0,0}

\hypersetup{
    bookmarks=true,         
    bookmarksnumbered=true,
    unicode=false,          
    pdftoolbar=true,        
    pdfmenubar=true,        
    pdffitwindow=false,     
    pdfstartview={FitH},    
    pdfauthor={Daisuke IIZASA},     
    colorlinks=true,       
    linkcolor=RoyalRed,      
    citecolor = RoyalRed,          
    linkbordercolor={1 0 0},
    urlcolor = RoyalRed, 
}
\begin{document}

\title{Enhanced longevity of the spin helix in low-symmetry quantum wells}

\author{Daisuke Iizasa}
\affiliation{Department of Materials Science, Tohoku University, Sendai 980--8579, Japan}
\affiliation{School of Chemical and Physical Sciences and MacDiarmid Institute for Advanced Materials and Nanotechnology, Victoria University of Wellington, P.O. Box 600, Wellington 6140, New Zealand}
\author{Makoto Kohda}
\affiliation{Department of Materials Science, Tohoku University, Sendai 980--8579, Japan}
\affiliation{Center for Spintronics Research Network, Tohoku University, Sendai 980--8577, Japan}
\affiliation{Center for Science and Innovation in Spintronics (Core Research Cluster), Tohoku University, Sendai 980--8577, Japan}
\affiliation{Division for the Establishment of Frontier Sciences, Tohoku University, Sendai 980--8577, Japan}
\author{Ulrich Z\umlaut{u}licke}
\affiliation{School of Chemical and Physical Sciences and MacDiarmid Institute for Advanced Materials and Nanotechnology, Victoria University of Wellington, P.O. Box 600, Wellington 6140, New Zealand}
\author{Junsaku Nitta}
\affiliation{Department of Materials Science, Tohoku University, Sendai 980--8579, Japan}
\affiliation{Center for Spintronics Research Network, Tohoku University, Sendai 980--8577, Japan}
\affiliation{Center for Science and Innovation in Spintronics (Core Research Cluster), Tohoku University, Sendai 980--8577, Japan}
\author{Michael Kammermeier}\email{michael.kammermeier@vuw.ac.nz}
\affiliation{School of Chemical and Physical Sciences and MacDiarmid Institute for Advanced Materials and Nanotechnology, Victoria University of Wellington, P.O. Box 600, Wellington 6140, New Zealand}

\begin{abstract}
In a semiconductor, collective excitations of spin textures usually decay rather fast due to D'yakonov-Perel' spin relaxation.
The latter arises from spin-orbit coupling, which induces 
wave-vector $(\beq{k})$ dependent spin rotations that, in conjunction with random disorder scattering, generate spin decoherence.
However, symmetries occurring under certain conditions can prevent the relaxation of particular homogeneous and inhomogeneous spin textures.
The inhomogeneous spin texture, referred to as  a persistent spin helix, is especially appealing as it enables us to manipulate the spin orientation while retaining a long spin lifetime.
Recently, it was predicted that such symmetries can be realized in zinc-blende two-dimensional electron gases if at least two growth-direction Miller indices agree in modulus and the coefficients of the Rashba and $\beq{k}$-linear Dresselhaus spin-orbit couplings are suitably matched [Kammermeier \textit{et al.}, \hyperlink{https://journals.aps.org/prl/abstract/10.1103/PhysRevLett.117.236801}{Phys. Rev. Lett. {\bf 117}, 236801 (2016)}].
In the present paper, we systematically analyze the impact of the symmetry-breaking $\beq{k}$-cubic Dresselhaus spin-orbit coupling, which generically coexists in these systems, on the stability of the emerging spin helices with respect to the growth direction.
We find that, as an interplay between orientation and strength of the effective magnetic field induced by the $\beq{k}$-cubic Dresselhaus terms, the spin relaxation is weakest for a low-symmetry growth direction that can be well approximated by a [225] lattice vector.
These quantum wells yield a 30\% spin-helix lifetime enhancement compared to  [001]-oriented electron gases and, remarkably, require a negligible Rashba coefficient.
The rotation axis of the corresponding spin helix is only slightly tilted out of the quantum-well plane.
This makes the experimental study of the spin-helix dynamics readily accessible for conventional optical spin orientation measurements where spins are excited and detected along the quantum-well growth direction.
\end{abstract}

\date{\today}

\maketitle

\section{Introduction}
\begin{figure*}[htbp]
        \centering 
        \includegraphics[keepaspectratio, scale=.5]{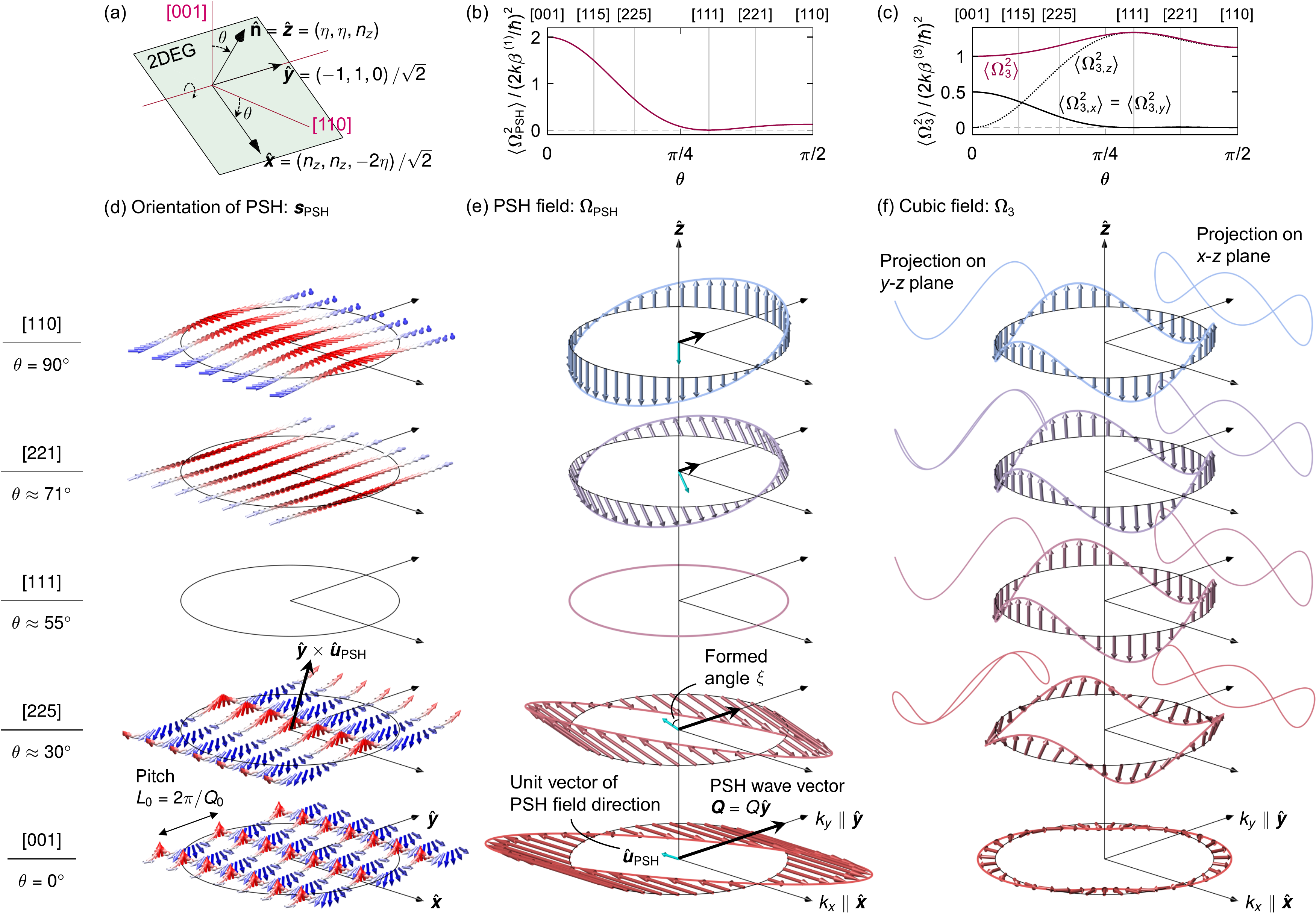}    
        \caption{(a) Illustration of the 2DEG-inherent coordinate system with respect to the crystal axes. The general growth direction $\hat{\bf n}$, that permits spin conservation, is characterized by the polar angle $\theta\in[0,\pi/2]$ in the [110]-[001] plane measured from the [001] axis. The basis is defined as $\hat{\beq{x}}=(n_z,n_z,-2\eta)/\sqrt{2}$,  $\hat{\beq{y}}=(-1,1,0)/\sqrt{2}$, and $\hat{\beq{z}}=\hat{\bf n}=(\eta,\eta,n_z)$, where $\eta=\sin\theta/\sqrt{2}$ and $n_z=\sqrt{1-2\eta^2}$. 
Both collinear SO field  $\psh$, Eq.~(\ref{soPSHfield}), for appropriately tuned Rashba and $\beq{k}$-linear  Dresselhaus SO coefficients, and $\beq{k}$-cubic Dresselhaus SO field $\3$, Eq.~(\ref{thirdangular}), are depicted in $\beq{k}$-space for specific growth directions in (e) and (f), respectively.
The mean squares $\langle\psh^2\rangle$ and $\langle\3^2\rangle$, averaged over all angles $\varphi$ of the in-plane wave vector $\bm{k}$, are shown in (b) and (c), respectively.
The orientation $\upsh$ of the PSH field, Eq.~(\ref{upsh}), is emphasized by the blue arrow in (e), which encloses the angle $\xi=\mathrm{arccos(\eta/\sqrt{2-3\eta^2})}$ with the surface normal $\hat{\bf n}$. 
The angle $\xi$  changes continuously from $\pi/2$  (in-plane) for [001] to 0 (out-of-plane) for [110]-oriented quantum wells. 
The $\theta$-dependent PSH wave vectors $\beq{Q}=Q \y$,  Eq.~(\ref{Qvalue}), which define the pitch of the PSH, Eq.~(\ref{pshorien}), are displayed as bold black arrows in (e).
The real space structure of the PSH $\bm{s}_{\rm PSH}$,  Eq.~\protect\refeq{pshorien}, determined by $\psh$ is illustrated in (d). The spin precession direction is reversed between [111] and [110] since $(\psh)_x$ switches sign.}
\label{one}
\end{figure*}

The quest to control electron spin in a solid has continued to date to advance progress in modern information technology~\cite{Wolf2001,Awschalom2007,Waldrop2016}. 
A series of potential future spintronic devices have been established, including spin transistors~\cite{Datta1990,Chuang2015}, all-spin logic gates \cite{Dery2007,BehinAein2010,Wen2016}, spin memories~\cite{Kikkawa1997,Kroutvar2004}, and spin lasers~\cite{Gothgen2008,Iba2011,Lee2014,FariaJunior2015,Lindemann2019}.
Especially for spin transistors, one usual fundamental requirement is the precise and reliable manipulation of spin orientation and lifetime.
In semiconductors, the spin-orbit (SO) coupling generates an effective magnetic field $\beq{\Omega}$, called SO field, which enables coherent control of the electron spin.
At the same time, the SO coupling induces the detrimental effect of spin decoherence via an efficient process known as D'yakonov-Perel' (DP) spin relaxation~\cite{Dyakonov1971a}. 
The origin of this effect lies in the wave-vector $(\beq{k})$ dependence  
of the SO field together with the presence of disorder.
Collisions of the spin carriers with impurities, phonons, or other carriers change the wave vector and thereby the spin precession axis uncontrolledly, which leads to randomization of the spins.
A way to overcome this problem is the realization of special symmetries that allow the emergence of persistent spin textures, as was found in electron and hole gases for appropriately tuned Rashba \cite{Rashba1960,Bychkov1984} and Dresselhaus \cite{Dresselhaus1955} SO strengths, strain, or curvature radius in tubular systems~\cite{Schliemann2003,Bernevig2006,Trushin2007,Sacksteder2014,
Dollinger2014,Wenk2016,Kammermeier2016,Kozulin2019,Kammermeier2020}.
In general, this symmetry becomes manifest in a SO field that is collinear in $\beq{k}$-space, and in spin-split circular Fermi contours $\epsilon_\pm$
that are related to each other by a shift of a constant wave vector $\pm\beq{Q}$, i.e., $\epsilon_-(\beq{k})=\epsilon_+(\beq{k}+\beq{Q})$~\cite{Bernevig2006}.
The collinearity of the SO field preserves any parallel-oriented homogeneous spin texture.
The second characteristic is associated 
with a new type of exact SU(2) spin-rotation symmetry of the Hamiltonian that allows for a full representation of the Lie algebra su(2)~\cite{Bernevig2006,Schliemann2017}.
It is fulfilled when the SO field
consists of first angular harmonics in the wave vector and ensures that the spins undergo a well-defined spin precession that is independent of the propagated path and, thus, robust against $\beq{k}$-randomizing disorder scattering~\cite{Schliemann2003,Wenk2016}.
This property allows the existence of an additional \textit{inhomogeneous} spin texture,
which due to their spiral structure is known as a \textit{persistent spin helix} (PSH)~\cite{Bernevig2006}.
As a decisive advantage over a homogeneous texture, the PSH facilitates a controllable spin precession over long distances.
It also entails numerous distinctive features in quantum transport that support experimental investigations~\cite{Sinitsyn2004,Shen2004,Schliemann2006,Badalyan2009,Kohda2012b,
Li2013,Schliemann2017,Kohda2017,Liu2020}.
\parag
In planar two-dimensional electron gases (2DEGs) with a zinc-blende structure, the existence of a PSH is well-established in quantum wells grown along the [001], [110], and [111] high-symmetry crystal axes~\cite{Zutic2004,Schliemann2017,Kohda2017}.
As illustrated in Fig.~\reffig{one}(e), the respective collinear SO field $\beq{\Omega}_{\rm PSH}$ for effectively $\beq{k}$-linear  SO couplings is either purely aligned with the 2DEG plane, out-of-plane, or vanishes completely.
Recently, it was predicted that a PSH can also be realized in low-symmetry growth directions provided that at least two Miller indices agree in modulus and the ratio of Rashba and effective $\beq{k}$-linear  Dresselhaus SO coefficients fullfills a certain relation~\cite{Kammermeier2016}.
Thereby, the angle between growth direction and collinear SO field can   be configured arbitrarily giving rise to new formations of a PSH \{cf. Fig.\,\reffig{one}(e) for the  exemplary SO field of [225] and [221]-oriented 2DEGs\}.
\parag
The stability of the PSH is, however, limited by an additional SO field arising from the $\beq{k}$-cubic Dresselhaus SO coupling that is generically present in these systems.
While its inclusion may not destroy the collinearity of the SO field, the presence of higher angular harmonics in the wave vector generally breaks the exact SU(2) spin-rotation symmetry of the Hamiltonian and causes a decay of the PSH.
Apart from this, it gives rise to new characteristic (in)homogeneous spin textures with extraordinary long lifetime, which for the sake of distinction we call long-lived, and the superior one among them is called \textit{longest}-lived spin textures.
In reciprocal space, the $\beq{k}$-cubic Dresselhaus SO field holds three-fold rotational symmetry, and its orientation and magnitude depend strongly on the growth direction as shown in Figs.\,\reffig{one}(c) and \reffig{one}(f).
As a consequence, the geometrical relations between the collinear and the symmetry-breaking part of the total SO field are complicated, and the induced relaxation effect is sensitive to the orientation of the quantum well.
Previous studies on the impact of the $\beq{k}$-cubic Dresselhaus SO field on the stability of the PSH were restricted to the well-established cases of [001] \cite{Liu2012,Koralek2009,Luffe2011,Luffe2013,Walser2012a,Kohda2012b,Salis2014,Kurosawa2015,Poshakinskiy2015,Ferreira2017}, [110] \cite{Ohno1999,Tarasenko2009,Iizasa2018}, and [111] \cite{Balocchi2011} quantum wells.
\parag
In this paper, we systematically explore the robustness of the PSH against the spin decoherence caused by the $\beq{k}$-cubic Dresselhaus SO field in zinc-blende 2DEGs of general crystal orientations.
The lifetime of the PSH is juxtaposed with that of the long-lived spin textures.
We complement a numerical Monte Carlo simulation of the random walk of collectively excited spins with an analysis of the spin diffusion equation to determine the PSH-lifetime-dependence on the growth direction.
The Monte Carlo approach resembles the experimental situation of a time-resolved magneto-optical Kerr-rotation microscopy that has been shown for some growth directions to be more suitable for the PSH-lifetime extraction than magneto-conductance measurements of weak antilocalization. 
The reason is that the latter characteristics are predominantly determined by the \textit{longest}-lived spin textures.
These often correspond to homogeneous 
spin textures whose extraordinarily long lifetimes prevent the emergence of the weak-antilocalization features necessary for reliable parameter fitting, as is the case, for instance, in [110] and [111] quantum wells~\cite{Hassenkam1997,Kammermeier2016,Iizasa2018}.
The supplementing analysis of the spin diffusion equation grants insight into the underlying physical mechanisms and allows us to derive an analytic expression for the general PSH lifetime and the special growth directions.
\parag
Our results reveal that the most robust PSH can be formed in quantum wells grown along a low-symmetry growth direction that is well approximated by a [225] lattice vector.
These systems yield a 30\% PSH lifetime enhancement compared to conventional [001]-oriented 2DEGs and require a negligible Rashba coefficient, allowing the realization of the most stable PSH in nearly symmetric quantum wells.
The origin of the suppressed spin relaxation along this direction is traced back to the strength and orientation of the $\beq{k}$-cubic Dresselhaus SO field. 
For the strength, a growth direction is favorable where the magnitude of the $\beq{k}$-cubic Dresselhaus SO coupling is reduced.
For the orientation, it is shown that, in an optimal configuration, the $\beq{k}$-cubic Dresselhaus SO field is perpendicular to the collinear SO field, and therewith lies in the rotation plane of the PSH.
In this case, the $\beq{k}$-cubic Dresselhaus SO field has globally the largest parallel component to the spin orientation of the PSH, which minimizes the relaxation.
Uniting both properties renders a growth direction ideal that is close to  the [225] axis.
Since the PSH rotation axis for the [225] quantum wells is only weakly tilted out of the 2DEG plane, the PSH can be observed in conventional optical spin orientation measurements where spins are excited and detected along the growth direction.
As a further advantage, we find that the lifetime of the long-lived homogeneous spin textures is particularly short in these systems, which supports the likelihood of observing weak antilocalization characteristics in magnetotransport measurements.
Our findings provide an complete and comprehensive picture of the stability of the PSH and the lifetime of the long-lived spin textures in general growth directions.
We identify the longest achievable lifetime for a PSH in the presence of $\beq{k}$-cubic Dresselhaus SO couplings in 2DEGs.
\parag
This paper is organized as follows.
In Sec.~\refchap{sec:GenericPSH}, we introduce the general SO field for PSH hosting quantum wells.
In Sec.~\refchap{sec:cubic}, we first discuss the impact of the $\beq{k}$-cubic SO field, and then we investigate its effect on the robustness of the PSH using Monte Carlo simulation and the spin diffusion equation in Secs.~\refchap{sec:MonteCarlo} and \refchap{sec:sde}, respectively.
In Sec.~\refchap{sec:analytic}, we derive analytical expressions for the relaxation rate of the PSH as well as the homogeneous spin texture, and we discuss the critical origin of the suppressed PSH decay for [225] quantum wells.
Lastly, we explore the experimental accessibility of the PSH in Sec.~\refchap{sec:accessibility}, and we close with a conclusion in Sec.~\refchap{sec:conlusion}.

\section{Persistent spin helix in generic 2D electron gases}\label{sec:GenericPSH}
\subsection{Spin-orbit fields for general growth directions}\label{sec:SOfields}

The PSH emerges under the precondition that at least two growth-direction Miller indices agree in modulus~\cite{Kammermeier2016}.
Without loss of generality, we focus on a general growth direction given by the unit vector $\hat{\bf n}$ lying in the first quadrant of the [110]-[001] crystal plane, i.e., $\hat{\bf n}=(\sin\theta/\sqrt{2},\sin\theta/\sqrt{2},\cos\theta)$. Here, the underlying basis vectors  point along the high-symmetry crystal directions [100], [010], and [001] and $\theta\in[0,\pi/2]$ denotes the polar angle measured from the [001] axis as shown in Fig.\,\reffig{one}(a).
For convenience and to adopt the notation of Ref.~\onlinecite{Kammermeier2016}, we introduce the parameter $\eta=\sin\theta/\sqrt{2}$, which implies that $n_z=\sqrt{1-2\eta^2}$, and we define new Cartesian basis vectors $\hat{\beq{x}}=(n_z,n_z,-2\eta)/\sqrt{2}$, $\hat{\beq{y}}=(-1,1,0)/\sqrt{2}$, and $\hat{\beq{z}}\equiv\hat{\bf n}=(\eta,\eta,n_z)$.
In this representation, the $\hat{\beq{x}}$ and $\hat{\beq{y}}$ axes span the conduction plane of the 2DEG, while the $\hat{\beq{z}}$ axis corresponds to the quantum-well growth direction.
\parag
In the vicinity of the $\Gamma$-point, the 2DEG is described by the Hamiltonian
\begin{equation}
\mathcal{H}=\cfrac{\hbar^2k^2}{2m}+\cfrac{\hbar}{2}\,(\beq{\Omega}_1+\3)\cdot\beq{\sigma}
\label{Hamiltonian}
\end{equation}
with effective electron mass $m$, in-plane wave vector $\beq{k}=(k_x,k_y)$,
and the vector of Pauli matrices $\beq{\sigma}=(\sigma_x,\sigma_y,\sigma_z)$.
The SO fields for the 2DEG-inherent coordinate system are derived from a general expression of the 2D-confined SO Hamiltonian as shown in Ref.~\onlinecite{Kammermeier2016} and briefly outlined in Appendix~\ref{app:SOF}.
The SO field contributions
\begin{align}
{\bf\Omega}_1&=\cfrac{2k}{\hbar}
\begin{pmatrix}
\left[\alpha+\beta^{(1)}(1+3\eta^2)n_z\right]\sin\varphi\\
\left[-\alpha+\beta^{(1)}(1-9\eta^2)n_z\right]\cos\varphi\\
-\sqrt{2}\beta^{(1)}\eta(1-3\eta^2)\sin\varphi
\end{pmatrix}
,\label{firstangular}
\end{align}
and
\begin{align}
\3&=\cfrac{2k}{\hbar}\beta^{(3)}
\begin{pmatrix}
(1-3\eta^2)n_z\sin3\varphi\\
-(1-3\eta^2)n_z\cos3\varphi\\
3\sqrt{2}\eta(1-\eta^2)\sin3\varphi
\end{pmatrix}
\label{thirdangular}
\end{align}
are sorted in terms of first and third angular harmonics in the in-plane wave vector, which is represented in polar coordinates, i.e., $k_x=k\cos\varphi$ and $k_y=k\sin\varphi$, with in-plane polar angle $\varphi$.
Here, the first angular harmonic contribution $\beq{\Omega}_1$ involves the Rashba and effective $\beq{k}$-linear  Dresselhaus SO coefficients  $\alpha=\gamma_{\rm R}\langle\mathcal{E}_z\rangle$ and $\beta^{(1)}=\gamma_{\rm D}(\langle k_z^2\rangle - k^2/4)$, respectively. 
Both coefficients scale with the material-specific bulk parameters $\gamma_{\rm R}$ and $\gamma_{\rm D}$ and constitute an average over the ground-state wave function determined by a self-consistently calculated confinement potential.
Thus, any inhomogeneities,  e.g., due to local fluctuations of the doping ions at the sides of the quantum well~\cite{Sherman2003a,Sherman2003b}, are disregarded.
The Rashba SO coupling is characterized by an electric field $\beq{\mathcal{E}}=\mathcal{E}_z\hat{\beq{z}}$ originating from a potential gradient along the growth direction $\hat{\beq{z}}$. 
The effective $\beq{k}$-linear  Dresselhaus SO coefficient is predominantly determined by the width and structure of the quantum well through the expectation value $\langle k_z^2\rangle$.
For instance, an infinite square-well potential of width $a$ yields $\langle k_z^2\rangle=(\pi/a)^2$.
Aside from this, the coefficient includes a small term $\propto k^2$ resulting from the first angular harmonic part of the $\beq{k}$-cubic Dresselhaus SO field. 
The third angular harmonic $\beq{k}$-cubic Dresselhaus contribution $\beq{\Omega}_3$ is distinguished by the prefactor $\beta^{(3)}=\gamma_{\rm D}k^2/4$.
Due to the proportionality $\propto k^2$, both Dresselhaus coefficients depend on the carrier sheet density $n_s$, i.e., at zero temperature $k$ is evaluated at the Fermi wave vector $k_{\rm F}=\sqrt{2\pi n_s}$.
For comparison, it is practical to work with the ratios of the SO coefficients where we employ the definitions $\Gamma_1=\alpha/\beta^{(1)}$ and $\Gamma_3=\beta^{(3)}/\beta^{(1)}$ hereafter.

\subsection{Collinear spin-orbit field and emergent persistent spin textures}\label{sec:PSHfield}

A vanishing of the $\beq{k}$-cubic SO contribution, i.e.,  $\Gamma_3=0$, together with an optimal ratio of the $\beq{k}$-linear  SO coefficients, i.e., $\Gamma_1=\Gamma_0:=(1-9\eta^2)n_z$, ensures a SO  field $\psh
:=\beq{\Omega}_1(\alpha= \beta^{(1)} \Gamma_0)$, in the following denoted as PSH field, that is collinear in $\beq{k}$-space and reads as~\cite{Kammermeier2016}
\begin{align}
\psh
&=\cfrac{2k}{\hbar}\beta^{(1)}
\begin{pmatrix}
2(1-3\eta^2)n_z\\
0\\
-\sqrt{2}\eta(1-3\eta^2)
\end{pmatrix}
\sin\varphi.
\label{soPSHfield}
\end{align}
As depicted in Figs.~\reffig{one}(b) and \reffig{one}(e), both the orientation and magnitude of $\psh$ alter with the growth direction.
The field is generally oriented  perpendicular to the $\y$ axis and encloses the angle $\xi=\mathrm{arccos(\eta/\sqrt{2-3\eta^2})}$ with the $\hat{\beq{z}} (\hat{\bf n})$ axis. 
Thereby, it allows a continuous modulation from an in-plane configuration for [001] to an out-of-plane configuration for [110] quantum wells.
It generally vanishes  for $\beq{k}\parallel\hat{\beq{x}}$ and maximizes for $\beq{k}\parallel\hat{\beq{y}}$ where the corresponding strength $\| \psh \|_{\rm max}$ is largest for $\eta=0$, which corresponds to a [001] growth direction.
In the special situation of a [111] 2DEG, i.e., $\eta=1/\sqrt{3}$, $\psh$ vanishes completely.
\parag
The PSH field leads to an SU(2) spin-rotation symmetry of the Hamiltonian, Eq.~(\ref{Hamiltonian}), that remains intact in the presence of spin-independent disorder and interactions~\cite{Bernevig2006,Schliemann2017}.
Considering a general spin density ${\beq{s}}(\beq{r},t)$ in real space with position vector $\beq{r}$ and time $t$, here and in the following locally and initially normalized as $\|{\beq{s}}(\beq{r},0)\|=1$, the SU(2) symmetry gives rise to two kinds of {persistent spin textures}: 
(i) A homogeneous spin texture ${\beq{s}}_{\rm homo}=\pm\upsh$ that is collinear with the direction of the PSH field
\begin{equation}
\upsh=\cfrac{{\rm sgn}(1-3\eta^2)}{\sqrt{2-3\eta^2}}
\begin{pmatrix}
-\sqrt{2}n_z\\
0\\
\eta
\end{pmatrix},
\label{upsh}
\end{equation}
which we define here as the unit vector of $\psh(k_y<0)$ [Fig.\,\reffig{one}(e)]. 
(ii) The PSH
\begin{align}
{\beq{s}}_{\rm PSH}(\beq{r})=&(\y\times\upsh)\cos({\beq{Q}}\cdot{\beq{r}}) - \y\sin({\beq{Q}}\cdot{\beq{r}}), \label{pshorien}
\end{align}
which spatially precesses about the $\upsh$ orientation and along the direction of $\pm\y$ ($\varphi=\pm\pi/2$) [Fig.~\reffig{one}(d)]. 
Here, we neglected an arbitrary phase shift for simplicity.
The sign function in Eq.~\refeq{upsh} implies an inversion of the precession axis of ${\beq{s}}_{\rm PSH}$ at [111] due to the sign switching of $(\psh)_x$ [cf. Figs.~\reffig{one}(e) and \reffig{one}(d)].
The pitch of the helix is defined by the PSH wave vector $\beq{Q}=Q\hat{\beq{y}}$ characterized by the maximum strength $\| \psh\|_{\rm max}$, i.e., \cite{Kammermeier2016}
 \begin{equation}
Q(\eta)=\frac{m}{\hbar k} \| \psh \|_{\rm max}=\Qo\sqrt{1-3\eta^2/2}|1-3\eta^2|, \label{Qvalue}
\end{equation}
where $Q(0):=\Qo = 4m\beta^{(1)}/\hbar^2$ represents the PSH wave-vector amplitude for a [001] 2DEG. 
We define the spin precession length $L$ as the (minimal) spatial length of one precession cycle, i.e., $L(\eta)=2\pi/Q(\eta)$, as illustrated in Fig.\,\reffig{one}(d) for $L(0):=L_0$.
The PSH wave vector is displayed as black bold arrows sketched in Fig.~\reffig{one}(e).
In accordance with the dependence of $\| \psh \|_{\rm max}$ on $\eta$, the magnitude of the PSH wave vector continuously decreases from the global maximum at [001] ($\eta=0$) until it vanishes at [111] ($\eta=1/\sqrt{3}$), and then it increases again until a local maximum is recovered at [110] ($\eta=1/\sqrt{2}$).

\section{Stability of the spin helix}\label{sec:stability}

\subsection{Impact of cubic Dresselhaus field}\label{sec:cubic}

Taking into account the $\beq{k}$-cubic Dresselhaus SO field $\3$ that involves the third angular harmonics in the wave vector, the PSH $\beq{s}_{\rm PSH}$ acquires a relaxation factor $\exp(-t/\tau_{\rm PSH})$ due to the violation of the exact SU(2) spin-rotation symmetry of the Hamiltonian.
The consequential finite PSH lifetime $\tau_{\rm PSH}$ depends on the strength and structure of $\3$ and its non-trivial geometric relations to the PSH field $\psh$.
Notably, 
the breaking of SU(2) symmetry is not necessarily accompanied by a destruction of the collinearity of the SO field but can be solely due to the presence of third angular harmonics in the wave vector.
Thus, the inclusion of $\3$ may continue to allow for a homogeneous spin texture with infinite lifetime but perhaps distinct orientation.
As will be discussed in more detail in Sec.~\ref{sec:analytic}, the underlying reason is that, in general, neither the PSH nor the homogeneous persistent spin texture, as defined in the previous section, are eigenstates of the spin diffusion equation any longer.
The eigenstates of the spin diffusion equation with the total SO field $\psh+\3$ give rise to new characteristic spin textures with particularly long spin lifetimes, which can have a different real-space structure.
These spin textures are, in the following, referred to as \textit{long-lived homogenous} and \textit{long-lived helical spin textures} depending on whether or not their spin orientation modulates in real space.
\parag
As highlighted in Fig.\,\reffig{one}(f), the SO field $\3$ holds three-fold  rotational symmetry, and its orientation and magnitude exhibit rich variations with the growth direction.
For a clearer understanding, we display in Fig.\,\reffig{one}(c) the $\theta$-dependence of the components of $\3$ together with the squared magnitude averaged over the directions of the in-plane wave vector, i.e., $\langle \3^2\rangle=\int_0^{2\pi}\3^2\,{\rm d}\varphi/(2\pi)$.
While for small angles of $\theta$ the in-plane components dominate, they become insignificant for large angles. 
The SO field is oriented purely in-plane for [001] and purely out-of-plane for [111] and [110] growth directions.
The latter two directions are special since $\3$ and therewith the total SO field is collinear and, hence, a homogeneously $\hat{\bm{z}}$-polarized spin texture does not decay.
The  mean square $\langle \3^2\rangle$ shows only weak modulations with $\theta$ where a global minimum (maximum) is obtained for [001] ([111]) quantum wells.
\parag
Hence, the individual modulations of the PSH field $\psh$ and the $\beq{k}$-cubic Dresselhaus SO field $\3$ yield an intricate dependence of the robustness of the persistent spin textures on the growth direction, which will be elaborated on in detail below.

\begin{figure*}
        \centering
        \includegraphics[keepaspectratio, scale=0.5]{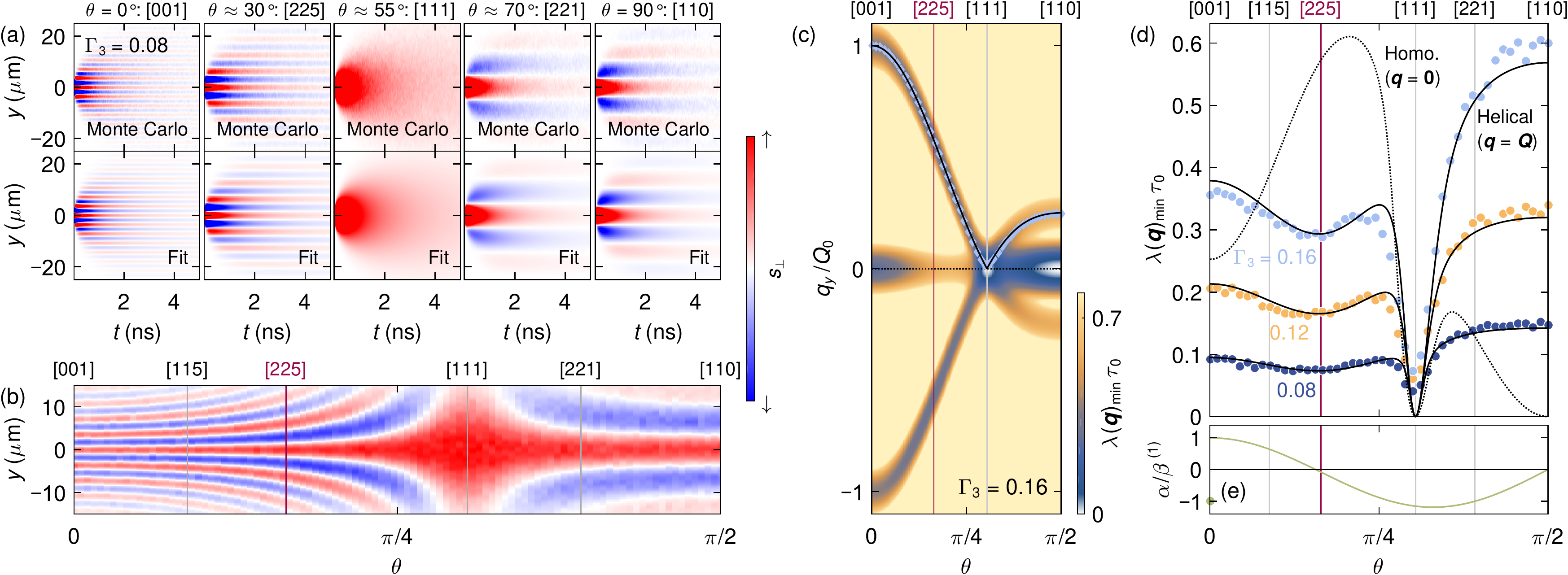}    
        \caption{(a) Monte-Carlo-simulated spatiotemporal evolution of the spin polarization $s_\perp$ along the $\y$ axis is shown in the first row for selected growth directions and a relative strength $\Gamma_3=0.08$ of the $\beq{k}$-cubic field. 
Reconstructions of the simulated $s_\perp$ by using the fit function Eq.\,\protect\refeq{sperp} are depicted in the second row. 
(b) Snapshot of $s_\perp$ along $\y$ at time $t=\SI{1}{ns}$ for continuous variations of the growth angle $\theta$.
(c) Map of the minima of the eigenvalues $\lambda_n(\beq{q})$ of the spin diffusion operator $\sdo$ in dependence of $\bm{q}=(q_x=0,q_y)$ and $\theta$ for $\Gamma_3=0.16$ in terms of the spin precession rate $1/\tau_0=D_sQ_0^2/(4\pi^2)$ for [001] quantum wells.
Black solid and dashed lines indicate the slices of $\beq{q}=\beq{Q}$ and $\beq{q}=\beq{0}$, respectively.
Extracted values of $Q$ from Monte Carlo simulations for $\Gamma_3=0.16$ using Eq.~\protect\refeq{sperp} are displayed as light blue diamonds.
(d) Computed eigenvalues (black solids) gathered along $\beq{q}=\beq{Q}$ in comparison with the extracted spin relaxation rate from Monte Carlo simulation (colored circles) for several $\Gamma_3$ values. The eigenvalue for $\beq{q}=\beq{0}$ is shown only at $\Gamma_3=0.16$ as a black dashed line.
 (e) PSH  condition $\alpha/\beta^{(1)}=\Gamma_0$ as a function of $\theta$.}
\label{two}
\end{figure*}

\subsection{Numerical Monte Carlo simulation}\label{sec:MonteCarlo}

To explore the relaxation of the PSH, we first conduct a numerical Monte Carlo simulation of the spin-random walk in a disordered system~\cite{Kiselev2000}.
Considering zero temperature and electronic states centered at the Fermi energy, an ensemble of $\SI{5e4}{spins}$ ($\beq{S}$) is aligned perpendicular to $\beq{\Omega}_{\rm PSH}$ at the initial time ${t=0}$.
The states are uniformly distributed over the Fermi circle with approximately isotropic Fermi wave vector $k_{\rm F}=\sqrt{2\pi n_s}$, where we select a carrier sheet density of $n_s=\SI{1.7e15}{m^{-2}}$. 
The carriers undergo a quasi-classical random walk with a ballistic motion between scattering events, which are considered elastic, isotropic, uncorrelated, and spin-independent.
The time evolution is characterized by a mean elastic scattering time $\tau=2D_s/v_{\rm F}^2=\SI{1.88}{ps}$  with Fermi velocity $v_{\rm F}=\hbar k_{\rm F}/m$ corresponding to a 2D diffusion constant $D_s=\SI{0.03}{m^2/s}$ and an effective mass of GaAs $m=0.067\, m_0$, where $m_0$ denotes the bare electron mass.
In each time interval of the ballistic motion, the spins propagate with Fermi velocity  $v_{\rm F}$ while precessing about the SO field following the differential equation $(\partial /\partial t)\beq{S}=(\psh+\3)\times\beq{S}$.
After time $t\gg\tau$, we locally detect spin projections perpendicular to $\psh$ and thereby extract the spin density component $s_\perp(\beq{r},t)$.
In accordance with the typical experimental scenario of an optical spin excitation, we assume an initialized Gaussian spin distribution in real space centered at $\beq{r}=\bm{0}$ with sigma width $w=\SI{0.5}{\mu m}$ defining the laser spot size.
We fix the effective $\beq{k}$-linear  Dresselhaus SO coefficient $\beta^{(1)}=\SI{5.0}{meV\AA}$ throughout all simulations while the Rashba SO parameter $\alpha$ varies with $\eta$ $(\theta)$ according to the relation $\Gamma_0$.
Among all crystal orientations, this gives a minimal pitch of $L_0=\SI{3.58}{\mu m}$, obtained for [001] quantum wells, which is much larger than mean free path $\tau v_{\rm F}=\SI{0.339}{\mu m}$, ensuring the DP regime of all Monte Carlo simulated spin dynamics.
To clarify the impact of the symmetry-breaking SO field  $\3$, we modify
the $\beq{k}$-cubic Dresselhaus parameter $\beta^{(3)}$ independently of $\beta^{(1)}$ although this is difficult to achieve in experiment due to the mutual dependence on the carrier density.
\parag
The first row of Fig.~\reffig{two}(a) collects the Monte-Carlo-simulated time evolution of $s_{\perp}$ along the $\y$ axis for several growth directions distinguished by $\theta$ with influence of $\beq{k}$-cubic SO field $\3$ of relative strength $\Gamma_3=0.08$ ($\beta^{(3)}=\SI{0.4}{meV\AA}$).
The initialized Gaussian spin polarization evolves into a {helical texture} with distinct precession lengths $L$, which reflects the $\theta$-dependence of the wave vector $Q$, Eq.~\refeq{Qvalue}. 
To highlight the continuous changes of the spin precession length, we plot the spatial evolution of $s_{\perp}$ at time $t=\SI{1}{ns}$ in dependence of $\theta$ in Fig.\,\reffig{two}(b).
For the [111] orientation, the helical structure disappears, which is consistent with the vanishing of $\psh$ and $Q$.
\parag
To extract the wave vector $Q$ and the relaxation rate of the remnant helical spin density $1/\tau_{\rm hel}$, we fit the data of the Monte Carlo simulation using the function~\cite{Salis2014}
\begin{align} 
s_{\perp}=& \frac{w^2}{w^2 + 2D_st}\,\exp{\left[{-\frac{y^2+2w^2Q^2D_st}{2(w^2 + 2D_st)}}\right]}\nonumber\\
&\times \exp\left({-\cfrac{t}{\tau_{\rm hel}}}\right)\cos{\left( \frac{2D_st}{w^2 + 2D_st}{Q}\,y\right)}.  
\label{sperp}
\end{align}
Setting $1/\tau_{\rm hel}$ and $Q$ as a free parameters, we fit the data by Eq.~\refeq{sperp} and obtain good agreement with Monte Carlo simulation as shown in the second row of Fig.~\reffig{two}(a).
It is noteworthy that we also carried out marginal adjustments of $D_s$ in the fitting procedure to compensate for minor deviations arising from the input value ($\approx3$\%) due to slight fluctuations of $\tau$ in the numerical simulation.
\parag
In the subsequent sections, we discuss the $\theta$-dependence of the  extracted parameters and show explicitly that low-symmetry quantum wells near a [225] orientation constitute the ideal system to maximize the PSH lifetime and explain the physical origin.

\subsection{Spin diffusion equation}\label{sec:sde}

To take a closer look at the impact of the $\beq{k}$-cubic SO field on the PSH dynamics obtained by the Monte Carlo simulation and to elucidate the underlying physical mechanism, we study the spin decoherence using the spin diffusion equation.
Numerous papers were devoted to tracking the spatiotemporal evolution of a spin density in different parameter regimes using semiclassical~\cite{Malshukov2000,Schwab2006,Yang2010,Luffe2011} or diagrammatic~\cite{Burkov2004,Stanescu2007,Wenk2010,Liu2012,Poshakinskiy2015} approaches.
\parag	
In this paper, we concentrate on the low-energy regime with weak SO coupling and disorder at zero temperature.
Selecting the Fourier representation with small frequencies $\omega$ and in-plane wave vectors $\beq{q}$, $\beq{k}$, the dynamics of the Fourier-transformed spin density $\tilde{\beq{s}}(\beq{q},\omega)={\int {\rm d}r^2 \int{\rm d}t\,e^{i(\omega t-\beq{q}\cdot\beq{r} )}{\beq{s}}(\beq{r},t)}$ is governed by the diffusion equation~\cite{Wenk2010,wenkbook,Kammermeier2016}
\begin{equation}
\beq{0}=(D_s\beq{q}^2-i\omega+1/\hat{\beq{\tau}}_{\rm DP})\tilde{\beq{s}}(\beq{q},\omega)-\cfrac{2\hbar\tau }{im} \langle(\beq{k}\cdot\beq{q})\beq{\Omega}\rangle\times\tilde{\beq{s}}(\beq{q},\omega)
\label{eq:diffeq0}
\end{equation}
with the DP spin relaxation tensor $(1/\hat{\bm{\tau}}_{\rm DP})_{ij}={\tau(\langle\bm{\Omega}^2\rangle\delta_{ij}-\langle\Omega_i\Omega_j\rangle)}$.
The average $\langle.\rangle$ is performed over all polar angles $\varphi$ of the wave vector $\beq{k}$.
Notably, if we account for anisotropic scattering, the first and third angular harmonic SO fields involve distinct scattering times $\tau_1$ and $\tau_3$, respectively~\cite{Knap1996}.
The above equation is still valid in this case, though, if one replaces $\tau\rightarrow \tau_1$ and $\beta^{(3)}\rightarrow \beta^{(3)}\sqrt{\tau_3/\tau_1}$.
It is practical to rewrite Eq.~(\ref{eq:diffeq0}) in terms of a diffusion operator $\sdo$, which comprises all dynamical properties and yields
\begin{equation}
\beq{0}=\left[\sdo(\beq{q})-i\omega\right]\tilde{\beq{s}}(\beq{q},\omega). \label{diffeq}
\end{equation}
The eigenvalues $\lambda_n(\beq{q})$ ($n=1,2,3$) of $\sdo(\beq{q})$ 
describe spin relaxation rates of a system with arbitrary Rashba and Dresselhaus SO fields according to Eqs.~(\ref{firstangular}) and (\ref{thirdangular}).
The explicit expression of $\sdo$ is presented in App.~\refchap{app1}.
\parag
Fig.~\reffig{two}(c) shows the smallest of the three eigenvalues $\lambda(\beq{q})_{\rm min}$ as a function of the wave vector {$q_y$ $(q_x=0)$ and the growth angle $\theta$ under the PSH condition $\Gamma_1=\Gamma_0$ and for $\Gamma_3=0.16$.
As emphasized by the black dotted and solid lines in Fig.~\reffig{two}(c), the eigenvalues exhibit generally three minima, where one occurs at $\beq{q}=\beq{0}$  and the other two at finite $\beq{q}=\pm\beq{Q}$, whose magnitude, despite the $\beq{k}$-cubic SO terms, is perfectly described by Eq.~(\ref{Qvalue}).
The local minima refer to the long-lived spin textures whereas the global minimum defines the \textit{longest}-lived or superior spin texture, which, depending on $\beq{q}$, can be either homogeneous ($\beq{q}=\beq{0}$) or helical ($\beq{q}\neq\beq{0}$).
We also show that the values of $Q$ extracted by Monte Carlo simulation in terms of $Q_0$, the light blue diamonds in Fig.~\reffig{two}(c), agree well with the ideal functional behavior in Eq.\,\refeq{Qvalue}.
\parag
We first focus on the helical texture and plot the spin relaxation rate $\lambda(\beq{Q})_{\rm min}=1/\tau_{\rm hel}$ in dependence of $\theta$ for several values of $\Gamma_3$ (black solid lines)  together with the respective results obtained by the Monte Carlo simulation (colored circles) in Fig.~\reffig{two}(d).
To emphasize that the parameters are selected in the desired regime where the spin lifetime exceeds the spin precession time, the spin relaxation rate is displayed in units of $1/\tau_0=D_s\Qo^2/(4\pi^2)$ corresponding to the maximal spin precession rate which is obtained for [001] quantum well.
We find excellent agreement for all $\Gamma_3$ values between the two different approaches and see a rich dependence of $1/\tau_{\rm hel}$ on $\theta$.
\parag
Firstly, the [110] direction shows a less robust spin texture compared to [001], being consistent with previous calculation \cite{Iizasa2018}.
Secondly, the salient vanishing of $1/\tau_{\rm hel}$ at [111] results from the existence of a homogeneous persistent spin texture as $\psh=\beq{0}$ and $\3$ is collinear with the [111] axis.
Since both excited and detected spin textures contain a finite component parallel to [111], the extracted spin relaxation rate corresponds to the homogeneous texture and not to a helical one.
Similar argument holds in the vicinity of [111], where the wave vector $Q$ is negligible and the long-lived texture is basically homogeneous.
As a homogeneous texture lacks the ability for manipulable spin orientation, we shall not be interested in these growth directions.
Thirdly and most remarkably, another local minimum occurs near the [225] low-symmetry growth direction as emphasized by the colored grid line in Fig.\,\reffig{two}(d).}
Compared to conventional [001]-oriented 2DEGs, we find here a spin lifetime enhancement of 30\% while a helical spin texture is retained. 
Further attractiveness of [225] is that the Rashba coefficient $\alpha$ almost vanishes for the PSH condition $\Gamma_0$ as calculated in Fig.\,\reffig{two}(e).
It implies that a symmetric quantum well already exhibits a PSH without the need to tune $\alpha$ electrically~\cite{Nitta1997}. 
This reduces complications arising from a possible inhomogeneity of $\alpha$, which causes local imbalances of the ratio of $\alpha/\beta^{(1)}$ and constitutes an additional source of spin relaxation~\cite{Sherman2003a,Sherman2003b,Liu2006,Glazov2010,Bindel2016}.
The actual vanishing point of $\alpha$ is at $\eta=1/3$ corresponding to an irrational Miller index, but it is well approximated by a [225] direction~\cite{Kammermeier2016}.
\parag
We now turn to the spin relaxation rate $\lambda(\beq{0})_{\rm min}$ of the homogeneous texture, which is displayed in Fig.~\reffig{two}(d) as a black dotted line for  $\Gamma_3=0.16$.
The rate vanishes at [111] and [110] because the total SO field remains collinear.
If we compare with the pertaining helical rate, the lifetime of the helical texture is only clearly superior to that of the homogeneous texture in the vicinity of the [225] direction.
At [225] we generally  find $\lambda(\beq{0})_{\rm min}\approx 2\lambda(\beq{Q})_{\rm min}$ for arbitrary reasonable values of $\Gamma_3$.
This implies  that a [225]-oriented 2DEG is also suitable for an experimental spin lifetime extraction using magneto-conductance measurements of the weak antilocalization, as further discussed in Sec.~\ref{sec:accessibility}.
\parag
In the following, we analyze the growth-direction dependence of the long-lived spin textures in detail and elucidate the origin of the $\theta$-dependent robustness of the PSH.

\subsection{Analytic discussion}\label{sec:analytic}

\begin{figure*}
        \centering
        \includegraphics[keepaspectratio, scale=.5]{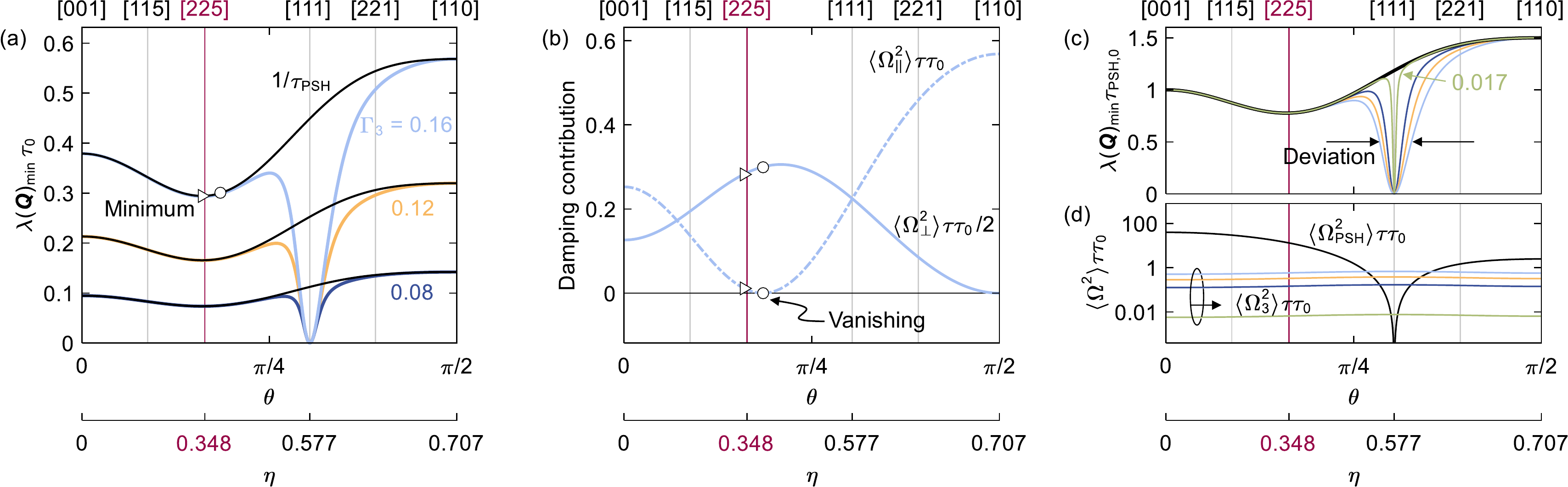}
        \caption{(a) Spin relaxation rate  of the long-lived helical spin texture $\lambda(\beq{Q})_{\rm min}=1/\tau_{\rm hel}$ 
        (colored solid lines) in comparison with the PSH relaxation rate $1/\tau_{\rm PSH}$ [Eq.~(\ref{analytical2})] (black solid lines) for several $\Gamma_3$ values in units of the spin precession rate in [001] quantum wells $1/\tau_0$.
The differences between the eigenvalue $\lambda(\beq{Q})_{\rm min}$ and $1/\tau_{\rm PSH}$ in the vicinity of [111]  implies that the long-lived helical spin texture deviates from the PSH.
A local minimum is found  at $\eta\approx0.341$, which is close to [225], where $\eta\approx0.348$ (red grid line). 
(b) Relaxation contributions parallel ($\langle \bo_\parallel^2\rangle\tau$) [Eq.~(\ref{flucpara})] and perpendicular ($\langle \bo_\perp^2\rangle\tau/2$) [Eq.~(\ref{flucperp})] to $\psh$ are plotted for $\Gamma_3=0.16$ as dashed and solid lines, respectively. 
The parallel contribution vanishes at $\eta\approx0.388$. 
In (a) and (b), the respective points $\eta\approx0.341$ and $\eta\approx0.388$ are highlighted as triangles and circles.
(c) Line shapes of $\lambda(\beq{Q})_{\rm min}$ and $1/\tau_{\rm PSH}$ are compared for several $\Gamma_3$ values in units of the PSH relaxation rate $1/\tau_{\rm psh,0}$ for [001] quantum wells. 
The smallest selected value $\Gamma_3=0.017$ corresponds to the experimentally extracted ratio in Ref.~[\onlinecite{Walser2012b}]. The range of growth angles where both rates deviate gets narrower as $\Gamma_3$ decreases.
(d) Mean squares of the SO strengths $\langle\psh^2\rangle$ and $\langle\3^2\rangle$ for different $\Gamma_3$ values with colors according to (c). 
The range of growth directions where $\langle\3^2\rangle$  exceeds $\langle\psh^2\rangle$ increases with $\Gamma_3$ yielding large deviations of the long-lived helical spin texture from the PSH and, thus, distinct relaxation rates.}
\label{three}
\end{figure*}

After including the $\beq{k}$-cubic SO field $\3$, the formerly persistent spin textures $\beq{s}_{\rm PSH}$ and $\beq{s}_{\rm homo}$, as defined in Sec.~\ref{sec:PSHfield}, are no longer eigenstates of the system. 
For this reason, the decay of these textures is in general not described by a single exponential function.
However since $\3$ usually constitutes a small correction to $\beq{\Omega}_1$, it is a good approximation to assume a single effective relaxation factor $\exp(-t/\tau_{\rm psh,homo})$ whose relaxation rate is given by projecting the diffusion operator $\sdo$ on these textures, i.e., $1/\tau_{\rm PSH}\equiv \langle\tilde{\beq{s}}_{\rm PSH}|\sdo (\beq{Q})|\tilde{\beq{s}}_{\rm PSH}\rangle$ and $1/\tau_{\rm homo}\equiv \langle\tilde{\beq{s}}_{\rm homo}|\sdo (\beq{0})|\tilde{\beq{s}}_{\rm homo}\rangle$.
Comparing these relaxation rates with those of the long-lived spin textures provides a deeper insight into the impact of the $\beq{k}$-cubic SO field.

\subsubsection{Relaxation of the spin helix}

\begin{figure}
        \centering
        \includegraphics[keepaspectratio, scale=0.5]{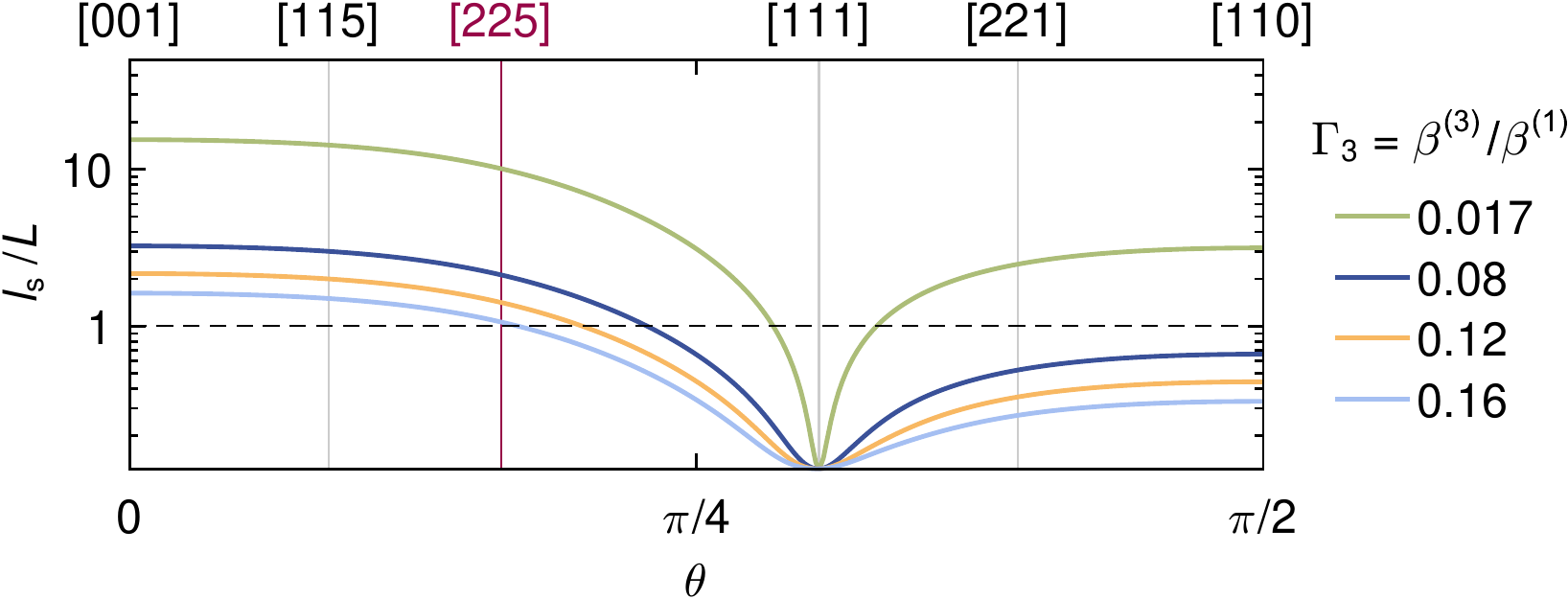}    
        \caption{Ratio between spin precession length $L=2\pi/Q$ and spin relaxation length of the long-lived helical spin texture $l_s=\sqrt{D_s}/\sqrt{\lambda(\beq{Q})_{\rm min}}$ is summarized for several $\Gamma_3$ values.}
\label{four}
\end{figure}

To reveal the underlying physical picture for the reduced relaxation in [225] quantum wells, however, it is more instructive to note that the contributions of $\3$ in the diffusion operator are decoupled from $\beq{q}$ and $\psh$.
This happens because the mixing of first and third angular harmonics in the wave vector averages to zero.
Thus, the relaxation is purely determined by the $\3$ terms in the DP tensor $1/\hat{\beq{\tau}}_{\rm DP}$ in Eq.~(\ref{eq:diffeq0}).
Recalling that in the DP formalism only perpendicular components of the SO field to the given spin orientation lead to a relaxation, we can apply this to the texture of the PSH and obtain
\begin{align}
\frac{1}{\tau_{\rm PSH}}&=\tau\int_0^{L}\frac{{\rm d}y}{L}\langle\left(\beq{\Omega}_3\times\beq{s}_{\rm PSH}\right)^2\rangle,
\label{integrals}
\end{align}
The integral represents the spatial average over a full spin precession of the PSH, e.g., $y\in[0,L]$.
This means that the component of the $\beq{k}$-cubic SO field parallel to $\beq{s}_{\rm PSH}$ does not contribute to relaxation.
According to this scenario, it is now practical to decompose $\3$ as
$\3=\bo_{\parallel}+\bo_{\perp}$, where $\bo_{\parallel}$ and $\bo_{\perp}$ are parallel and perpendicular to $\psh$, respectively.
Further decomposing $\bo_\perp=\bo_{\perp,1}+\bo_{\perp,2}$ with $\bo_{\perp,1}\parallel\y$ and $\bo_{\perp,2}\parallel(\y\times\upsh)$ simplifies Eq.~\refeq{integrals} and produces the analytical solution of the growth-dependent PSH relaxation rate, that is,
\begin{align}
\cfrac{1}{\tau_{\rm PSH}}&=  \langle \bo_{\parallel}^2\rangle \tau +\cfrac{\langle \bo_{\perp}^2\rangle\tau}{2} \label{analytical1},
\end{align}
which gives explicitely
\begin{align}
\cfrac{1}{\tau_{\rm PSH}}&=\cfrac{4\pi^2}{\tau_0}\cfrac{3 - 17 \eta^2 + 85 \eta^4 - 171 \eta^6 + 108 \eta^8}{8 - 12 \eta^2}\Gamma_3^2,\label{analytical2}
\end{align}
as a consequence of the relaxation contributions
\begin{align}
\langle\bo_\parallel^2\rangle\tau={}&\cfrac{4\pi^2}{\tau_0}\cfrac{(1 - 8 \eta^2 + 9 \eta^4)^2}{4-6\eta^2}\Gamma_3^2, \label{flucpara}\\
\cfrac{\langle \bo_{\perp}^2\rangle\tau}{2}={}&\cfrac{4\pi^2}{\tau_0}\cfrac{(1-\eta^2)(1+18\eta^2-27\eta^4)n_z^2}{8-12\eta^2}\Gamma_3^2. \label{flucperp}
\end{align}
As it becomes apparent from Eq.~\refeq{analytical1}, the parallel ($\bo_\parallel$) and perpendicular ($\bo_{\perp}$) components of $\3$ to $\psh$ affect the PSH relaxtion with different weighting, 1 and 1/2.
This results from the fact that $\bo_\parallel$ generates relaxation of the local spin orientation of the PSH over the full precession cycle as it is generally perpendicular to $\beq{s}_{\rm PSH}$, whereas $\bo_{\perp}$ is locally parallel to $\beq{s}_{\rm PSH}$, which partially protects the PSH from relaxation.
\parag
Figure \reffig{three}(a) compares the PSH relaxation rate $1/\tau_{\rm PSH}$ using the analytical expression Eq.~\refeq{analytical2} with the rate of the long-lived helical textures $1/\tau_{\rm hel}=\lambda(\beq{Q})_{\rm min}$ calculated with Eq.~\refeq{diffeq} for several $\Gamma_3$ values.
Aside from a narrow region in the vicinity of [111], the PSH relaxation rate matches well the long-lived helical rate for all selected $\Gamma_3$.
The close agreement indicates that the structure of the long-lived helical spin textures does not deviate much from the PSH.
We find that the local minimum of $1/\tau_{\rm PSH}\approx 0.78/\tau_{\rm psh,0}$, where $1/\tau_{\rm psh,0}=3\pi^2\Gamma_3^2/(2\tau_0)$ is the PSH relaxation rate for [001] quantum wells, emerges at $\eta=\sqrt{5-\sqrt{13}}/\sqrt{12}\approx0.341$ [cf. the triangle at the curve for $\Gamma_3=0.16$ in Fig.\,\reffig{three}(a)],  which is indeed close to [225], where $\eta=2/\sqrt{33}\approx 0.348$ (red grid lines in Fig.~\reffig{three}).
From the different weighting in Eq.~\refeq{analytical1}, it is reasonable to assume that this local minimum coincides with a minimum of the relaxation term $\langle\bo_\parallel^2\rangle\tau$, Eq.~\refeq{flucpara}, which means that $\3$ is perpendicular to $\psh$.
In  Fig.\,\reffig{three}(b), we display the individual relaxation contributions in terms of $1/\tau_0$, Eqs.\,\refeq{flucpara} and \refeq{flucperp}. 
The blue dashed line represents the contribution by $\bo_\parallel$, which drops to zero at $\eta=\sqrt{4-\sqrt{7}}/3\approx 0.388$ as indicated by a black circle.
Hence, the vanishing point of the parallel contribution at $\eta\approx 0.388$ does not perfectly agree with the suppressed relaxation rate as shown in Fig.\,\reffig{three}(a).
The reason is that, the magnitude of the perpendicular contribution varies simultaneously, as shown by the blue solid line in Fig.~\reffig{three}(b), and it holds a large magnitude at $\eta\approx 0.388$ (black circle). 
After normalizing the $\beq{k}$-cubic field $\3 \rightarrow \3 / \langle\|\3\|\rangle$, we find that the minimum occurs precisely at the expected value of $\eta\approx 0.341$, which confirms the previous assumption.
Consequently, the local suppression of the PSH relaxation rate is a combined effect of the interplay between the magnitude and orientation of $\3$.
Furthermore, the PSH relaxation rate becomes largest at [110] where $\langle\bo_\parallel^2\rangle\tau$ has a global maximum even though the perpendicular contribution vanishes since $\3$ is parallel to $\psh$.
\parag
We address now the role of the magnitude of $\3$ in the vicinity of growth directions where the spin relaxation rates of PSH and long-lived helical textures deviate, i.e., near [111]. 
Figure \reffig{three}(c) compares both relaxation rates, in units of $1/\tau_{\rm psh,0}$, for different values of $\Gamma_3$ ranging from  $0.017$ to $0.16$ [cf. Fig.~\reffig{three}(a)].
The former magnitude was experimentally obtained in Ref.~[\onlinecite{Walser2012b}].
We see that the growth-angle range where the line shapes of the relaxation rates of both texture types deviate becomes narrower as $\Gamma_3$ decreases.
The origin of the deviation is explained by the magnitude ratio between $\psh$ and $\3$.
Figure \reffig{three}(d) shows the average SO field magnitudes  $\langle\psh^2\rangle$ and $\langle\3^2\rangle$ in units of $1/(\tau \tau_0)$, where the plot colors are chosen in accordance with Fig.~\reffig{three}(c).
Since $\langle\psh^2\rangle$ rapidly drops down towards zero near [111], $\langle\3^2\rangle$ exceeds $\langle\psh^2\rangle$ and the SO field is dominated by $\3$.
Here, the PSH field is negligible compared to the $\beq{k}$-cubic SO field, implying that the eigenstate of diffusion operator strongly differs from the PSH.
The range of growth angles where $\3$ is dominant enlarges as $\Gamma_3$ increases. 
\parag
Since for practical applications spin functionalities must be implemented within the spin lifetime, it is desirable that $\Gamma_3$ is small enough that the spin precession length $L$ is much larger than the spin relaxation length of the long-lived helical spin texture $l_s=\sqrt{D_s\tau_{\rm hel}}=\sqrt{D_s}/\sqrt{\lambda(\beq{Q})_{\rm min}}$, i.e., $l_s / L \gg 1$.
The ratio $l_s / L$ is plotted in Fig.~\reffig{four}, which shows that for most growth directions a reasonable magnitude of $\Gamma_3$ is of the order of $10^{-2}$ or smaller.
As a concrete example for typical experimental values, we list $\tau_{\rm PSH}$ along other relevant quantities for several quantum wells with different growth directions in Tab.~\ref{tab:values}.
Consequently, for parameters of interest the growth-angle range where the rates of the PSH and the long-lived helical texture deviate is quite narrow [cf. Fig.~\ref{three}(c)], and Eq.~\refeq{analytical2} is a good solution for general growth directions.
Also, low-angle growth directions from [001] to approximately  [225] are less sensitive to the relaxation due to $\beq{k}$-cubic terms and, therefore, suitable candidates for applications.
Finally, it should be mentioned that comparing the spin precession lengths of the PSH in [225] and [001] quantum wells, the former is larger by approximately 50\%.

\subsubsection{Relaxation of the homogeneous spin texture}

\begin{figure}[t]
        \centering
        \includegraphics[keepaspectratio, scale=0.5]{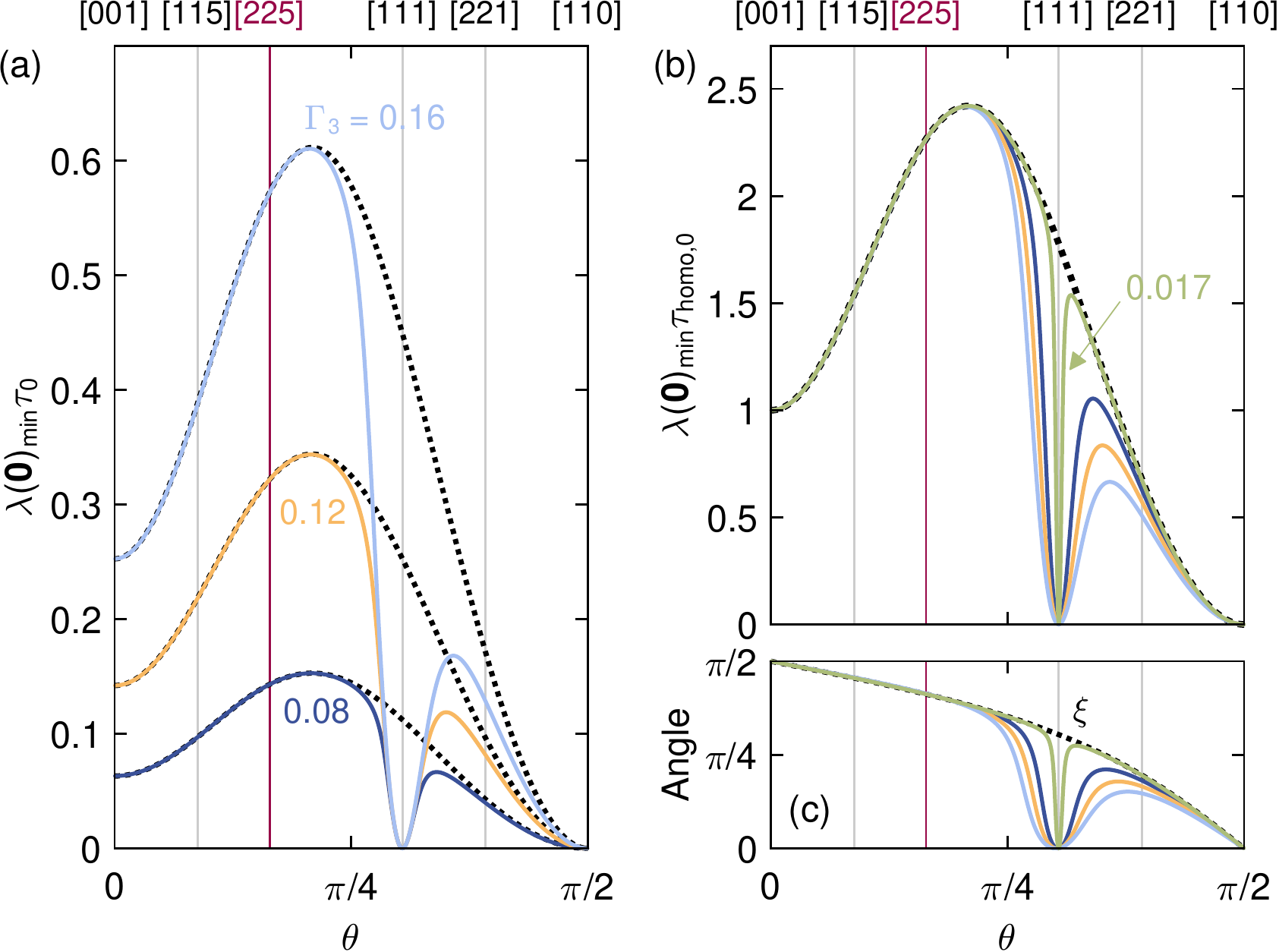}
        \caption{(a) Spin relaxation rates of the long-lived homogeneous texture $\lambda(\beq{0})_{\rm min}$ (colored solid lines)  in comparison with the relaxation rate $1/\tau_{\rm homo}$ [Eq.~(\ref{eq:relaxation_homo})] of the homogeneous spin texture $\beq{s}_{\rm homo}=\pm \upsh$ (black dashed line) for different values of $\Gamma_3$ in units of the spin precession rate in [001] quantum wells $1/\tau_0$.
The latter texture is collinear with the the PSH field orientation $\pm\upsh$ and, thus, persistent for $\3=\beq{0}$ (cf. Sec.~\ref{sec:GenericPSH}).
 (b)  Line shapes of $\lambda(\beq{0})_{\rm min}$ (colored solid lines) and $1/\tau_{\rm homo}$ (black dashed line) are shown for several $\Gamma_3$ values in units of the relaxation rate  $1/\tau_{\rm homo,0}$ for [001] quantum wells. 
  Analogously to the comparison of the long-lived helical texture with the PSH, Fig.~\ref{three}(c), 
  the range of growth angles where both relaxation rates deviate gets narrower as $\Gamma_3$ decreases.
  The smallest selected value $\Gamma_3=0.017$ corresponds to the experimentally extracted ratio in Ref.~[\onlinecite{Walser2012b}]. 
(c) Angle between the quantum-well growth direction and the spin orientation of the long-lived homogeneous spin texture (colored solid lines),  or  the homogeneous texture $\beq{s}_{\rm homo}$ (black dashed line), which is $\xi=\mathrm{arccos(\eta/\sqrt{2-3\eta^2})}$, for the analogous values of $\Gamma_3$ as in (b).}
\label{fig:homo}
\end{figure}

While the PSH plays a prominent role for the functionality of spin transistors, the dynamics of homogeneous spin textures is more relevant in other devices such as spin lasers~\cite{Gothgen2008,Iba2011,Lee2014,FariaJunior2015,Lindemann2019}.
For instance, a homogeneous spin texture is typically generated by optical spin orientation due to interband absorption of circularly polarized light when the illumination spot size exceeds the sample size~\cite{Dyakonov1971,Dyakonov1984}.
\parag
Thus, for a comprehensive understanding we investigate now the relaxation of the long-lived spin texture ${\beq{s}}_{\rm homo}=\pm\upsh$, which is homogeneously aligned parallel to the direction of the PSH field.
Following above arguments, its relaxation rate can be computed as $1/\tau_{\rm homo}=\tau\langle\left(\beq{\Omega}_3\times\upsh\right)^2\rangle$, which gives
%
\begin{align}
\frac{1}{\tau_{\rm homo}}&=\cfrac{4\pi^2}{\tau_0}\cfrac{n_z^2(1+17 \eta^2 -45 \eta^4 +27 \eta^6 )}{4 - 6 \eta^2}\Gamma_3^2.
\label{eq:relaxation_homo}
\end{align}
In Fig.~\ref{fig:homo}(a), the spin relaxation rate $1/\tau_{\rm homo}$ (black dashed lines) is displayed together with the rate of the long-lived homogeneous spin texture $\lambda(\beq{0})_{\rm min}$ (colored solid lines) for different values of $\Gamma_3$ in units of the spin precession rate in [001] quantum wells $1/\tau_0$.
Similarly to the PSH and the long-lived helical spin texture, we find good agreement between both relaxation rates, apart from a narrow region near [111].
The range of growth angles where both rates deviate becomes smaller as $\Gamma_3$ is reduced, which is depicted in Fig.~\ref{fig:homo}(b).
Typical values of $\tau_{\rm homo}$ for realistic parameter configurations are listed in Tab.~\ref{tab:values} for several quantum wells with distinct orientations.
For better comparison, the rates are rescaled in units of the relaxation rate $1/\tau_{\rm homo,0}=\pi^2\Gamma_3^2/\tau_0$ of ${\beq{s}}_{\rm homo}$ for [001] quantum wells.
In both figures, we notice that the relaxation rates have a pronounced global maximum, which we generically find to occur at $\eta\approx 0.431$ and which yields an increase by a factor of 2.4 compared with the rate for [001] quantum wells $1/\tau_{\rm homo,0}$.
For the [225] quantum wells, we find that the relaxation rate of the homogeneous mode is larger by a factor of 1.94 compared to the PSH relaxation rate $1/\tau_{\rm PSH}$, Eq~(\ref{analytical2}), at [225]. 
On the contrary, quantum wells with growth direction ranging from the vicinity of [111] to [110] facilitate very long spin lifetimes, with [111] and [110] offering persistent solutions.
Lastly, we plot in Fig.~\ref{fig:homo}(c), for the analogous values of $\Gamma_3$ as in Fig.~\ref{fig:homo}(b), the angle between the quantum-well growth direction and the spin orientation of the long-lived homogeneous spin texture (colored solid lines) and the homogeneous texture ${\beq{s}}_{\rm homo}$ (black dashed line), respectively.
The latter angle is given by the expression $\xi=\mathrm{arccos(\eta/\sqrt{2-3\eta^2})}$.
In the small region in the vicinity of [111], the spin orientation of the long-lived homogeneous spin texture becomes nearly parallel to the growth direction and strongly differs from ${\beq{s}}_{\rm homo}$, which explains the large discrepancy of the corresponding spin relaxation rates $1/\tau_{\rm homo}$ and $\lambda(\beq{0})_{\rm min}$.
\parag
As an implication for spintronic devices where long-lived homogeneous spin textures are desirable, e.g., for threshold reduction in spin lasers~\cite{Gothgen2008,Iba2011,Lee2014}, the [111] and [110] quantum wells are most appealing. Aside from the diverging spin lifetime, the corresponding spin polarization is perpendicular to the quantum-well plane ($\xi=0$), which often corresponds to the favorable excitation direction.

\newcommand{\ra}[1]{\renewcommand{\arraystretch}{#1}}
\ra{1.3}
\begin{table}
\centering
\caption{Realistic values in GaAs quantum wells for the relaxation times of the PSH $\tau_{\rm PSH}$ and the long-lived homogeneous spin texture $\tau_{\rm homo}$  using Eqs.~(\protect\ref{analytical2}) and (\protect\ref{eq:relaxation_homo}), respectively. 
We employ an effective mass $m=0.067m_0$,  a carrier sheet density $n_s=\SI{1.7e15}{m^{-2}}$, and spin diffusion constant $D_s=\SI{0.03}{m^2/s}$.
Accordingly, for a bulk Dresselhaus parameter $\gamma_{\rm D}=\SI{8.0}{eV\AA^3}$~\cite{Kohda2017}, the effective cubic Dresselhaus coefficient becomes $\beta^{(3)}=\gamma_{\rm D}k_{\rm F}^2/4=\gamma_{\rm D}\pi n_s/2=\SI{0.21}{meV\AA}$. 
Using a typical linear Dresselhaus coefficient $\beta^{(1)}=\SI{5.0}{meV\AA}$~\cite{Walser2012b} yields $\Gamma_3=\beta^{(3)}/\beta^{(1)}=0.042$.
The Rashba coefficient $\alpha$ and the spin precession length $L$ follow from the growth-direction-dependent relations $\alpha(\eta)=\Gamma_0(\eta)\beta^{(1)}$ and $L(\eta)=2\pi/Q(\eta)$ (cf. Sec.~\ref{sec:PSHfield}). }
\vspace{0.2cm}
\begin{ruledtabular}
\begin{tabular}{ccccc}
${\hat{\bf n}}$  &  $\tau_{\rm PSH}$ (ns)    &  $\tau_{\rm homo}$ (ns) &  $\alpha$ ($\SI{}{meV\AA}$)      &   $L$ ($\mathrm{\mu m}$)\\  
\hline
[001] & 15.8 & 23.7 & 5.0& 3.58\\\relax
[115] & 18.0  & 15.4 & 3.2& 4.14\\\relax
[225] & 20.3 & 10.5 & $-0.4$& 6.22\\\relax
[111] & - & $\infty$ & $-5.8$   & $\infty$\\\relax
[221] & 11.0 & 34.9 & $-5.0$& 18.6\\\relax
[110] & 10.5 & $\infty$ & 0 & 14.3 \\
\end{tabular}
\end{ruledtabular}
\label{tab:values}
\end{table}

\begin{figure}[t]
        \centering
        \includegraphics[keepaspectratio, scale=0.5]{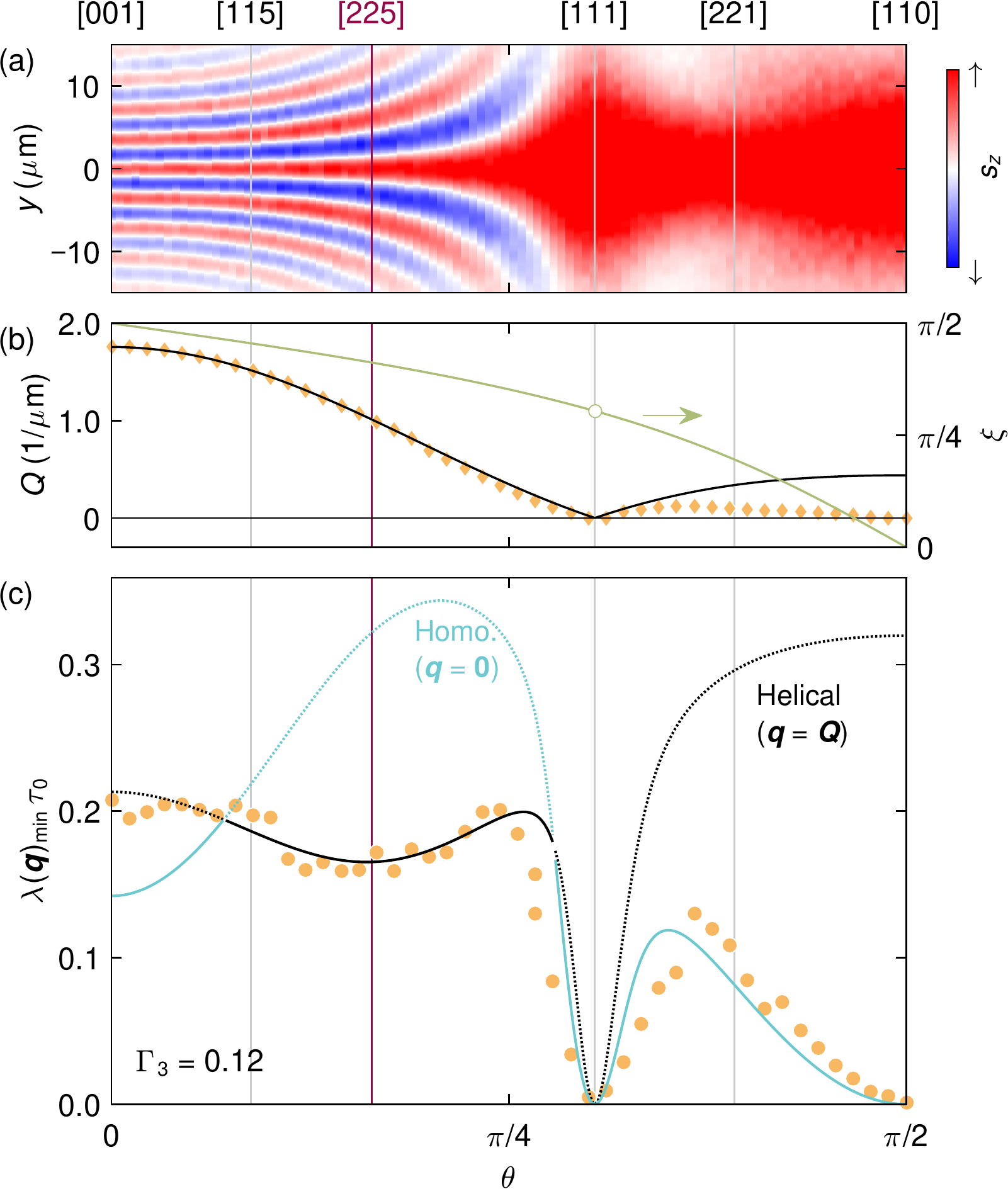}    
        \caption{
        (a) Monte-Carlo simulated component $s_z$ at time $t=\SI{1}{ns}$ of an initialized spin density $\beq{s}(t=0)\parallel \hat{\beq{z}}$ for $\Gamma_3=0.12$ in dependence of the growth angle $\theta$.
A helical spin texture emerges from [001] to approximately  [111], while a homogeneous texture is prevalent from [111] to [110]. 
(b)~Yellow diamonds represent the extracted wave vector $Q$ while black solid line is calculated by Eq.~(\ref{Qvalue}). 
Green solid line shows the angle $\xi$ between PSH field $\psh$ and surface normal~$\hat{\beq{z}}$.
(c)~Extracted spin relaxation rate is shown as yellow circles. Computed eigenvalues for homogeneous ($\lambda(\beq{0})_{\rm min}$) and helical ($\lambda(\beq{Q})_{\rm min}$) spin texture are displayed as blue and black dotted lines, respectively.
The global minimum of the relaxation rate corresponding to the \textit{longest}-lived spin textures is highlighted as solid line on the curves for the homogeneous and helical relaxation rates.}
\label{five}
\end{figure}
\subsection{Experimental accessibility}\label{sec:accessibility}

Finally, we discuss the accessibility of the PSH by optical experiments such as time-resolved Kerr-rotation microscopy, in which spins are typically excited and measured perpendicular to the quantum-well plane.
To obey this experimental restriction, we simulate the spatiotemporal evolution of the initial spin texture $\beq{s}(t=0)\parallel\hat{\beq{z}}$  using the Monte Carlo simulation as described in Sec.~\ref{sec:MonteCarlo}.
We employ the same parameter settings as before, where we restrict ourselves to the case of $\Gamma_3=0.12$.
In this new configuration, the spin excitation direction is tilted with respect to the direction of $\psh$ and therefore simultaneously polarizes long-lived  homogeneous and helical spin textures even for $\Gamma_3=0$.
The respective tilt angle depends on the quantum-well growth direction.
For the specific growth direction [001], the homogeneous texture is not excited as $\psh$ is completely in-plane~\cite{Salis2014} whereas we polarize only the homogeneous texture at [111] and [110] due to the configurations $\psh=0$, $\3\parallel \hat{\beq{z}}$ and $\psh\parallel \hat{\beq{z}}$, $\3\parallel \hat{\beq{z}}$, respectively.
\parag
Figure~\reffig{five}(a) shows the spatial distribution of $s_z$ along the $\y$ axis collected for various $\theta$ values at the slice of time $t=\SI{1}{ns}$.
The appearance of a helical texture for growth angles $\theta$ from 0 to approximately $\pi/4$ is obvious.
We attribute this to the large angle $\xi$ between $\psh$ and the $\hat{\beq{z}}$ axis [cf. Fig.~\reffig{five}(b)], which allows to excite the helical texture.
Using the fit function Eq.~\refeq{sperp}, we extract the wave vector $Q$ and the spin relaxation rate, which are displayed in Figs.~\reffig{five}(b) and \reffig{five}(c).
For angles $\theta<\pi/4$, we obtain a spin relaxation rate and wave vector $Q$ that agree well with the ideal values for the PSH [cf. Eqs.~(\ref{Qvalue}) and (\ref{analytical2})].
Therefore, we can readily access the PSH for different directions including the superior lifetime direction [225] in conventional optical measurements and take advantage of manipulable spin orientation.
For larger angles $\theta$ corresponding to growth directions from [111] to [110], the angle $\xi$ further decreases and the extracted relaxation rates and wave vectors belong to the homogeneous spin textures, which have the superior lifetime along these directions.
Also, the spin precession length of the helical textures becomes very long, which makes it difficult to distinguish from the homogeneous texture.
Apart from that, the magnitude of the PSH field becomes insignificant in comparison to the $\beq{k}$-cubic SO field, which yields large deviations of the long-lived helical spin texture from the PSH and makes the fit function Eq.~\refeq{sperp} unsuitable.
\parag
Finally, we look at the prospects to study the PSH lifetime limitation in magnetoconductance measurements of the weak antilocalization.
The characteristic weak-antilocalization feature, namely the position of the magnetoconductance minima, which is necessary for a reliable parameter fitting, is predominantly determined by the \textit{longest}-lived spin texture~\cite{Faniel2011,Yoshizumi2016,Kammermeier2016}.
In particular, if its lifetime is much longer than the electron-dephasing time, the minima disappear and only weak-localization features are seen.
Hence, to extract the PSH lifetime and the related $\beq{k}$-cubic Dresselhaus SO coefficient, it is desirable that the minima are still observable at the optimal ratio of $\beq{k}$-linear SO coefficients  $\alpha/\beta^{(1)}=\Gamma_0$, where for $\3=0$ persistent spin textures appear, and that the helical spin texture is superior.
The relaxation rate of the \textit{longest}-lived spin texture  for general growth directions is highlighted by the solid line in Fig.\,\reffig{five}(c) where the black and blue color corresponds to the helical and homogeneous spin texture, respectively.
We see that the homogeneous texture shows clear dominance for a wide range of growth directions.
The lifetime discrepancy between helical and homogeneous textures is most pronounced in [110] quantum wells, where the homogeneous rate vanishes while the helical rate reaches a maximum.
For this reason, we only observe weak localization even for a large $\beq{k}$-cubic SO field in [110] quantum wells due to the presence of persistent homogenous spin textures~\cite{Iizasa2018,Hassenkam1997}. 
Here, observing a crossover to weak antilocalization requires the breaking of the PSH condition $\Gamma_0$ via an increasing Rashba term~\cite{Hassenkam1997}.
Similarly, we can expect the extraordinary small relaxation rates of the homogeneous spin textures for growth directions between [111] and [110] to prevent the emergence of the weak antilocalization features. 
Intriguingly, the homogeneous relaxation rate far exceeds the helical rate
in the vicinity of the growth direction [225].
The large separation of the two relaxation branches implies that such quantum wells are suitable to extract the PSH lifetime as well as the $\beq{k}$-cubic Dresselhaus SO strength in magnetotransport measurements.
Also, the required near absence of the Rashba coefficient in the PSH case in [225] simplifies the parameter fitting process.
\parag
To sum up, [225] quantum wells are not only good candidates to enhance the PSH lifetime but also represent an excellent platform for a comparative study of the spin relaxation in optical and magnetotransport measurements.

\section{Conclusion}\label{sec:conlusion}

We have investigated the lifetime limitations of the PSHs due to the presence of third angular harmonics in the $\beq{k}$-cubic Dresselhaus SO field in 2DEGs of general growth directions.
A numerical approach using Monte Carlo simulations in conjunction with an analysis of the spin diffusion equation provides detailed knowledge of  the robustness of the long-lived spin textures.
\parag
Our findings reveal that a crystal orientation where the $\beq{k}$-cubic SO field is perpendicular to the collinear SO field suppresses the decay of the PSH because it partially protects local spin orientations from relaxation.
In combination with additional small modulations of the $\beq{k}$-cubic SO field magnitude, it is shown that the most robust PSH is formed in a low-symmetry quantum well whose orientation is well approximated by a [225] lattice vector.
Remarkably, the realization of a PSH in such a system requires a nearly vanishing Rashba SO coefficient.
This enables the utilization of a symmetric quantum well which mitigates complications that are connected to the electrical gate-tuning or a  spatial inhomogeneity of the Rashba SO strength.
We demonstrate explicitely that the PSH in [225] quantum wells can be experimentally accessed by optical spin excitation/detection measurements. 
Additionally, we point out that the PSH lifetime clearly exceeds the one of the long-lived homogeneous spin textures which makes it also a suitable candidate for an experimental extraction via weak (anti)localization measurements.
\parag
The results provide a complete picture of the stability of the PSH and the long-lived spin textures in general growth directions and define the longest possible PSH lifetime in 2DEGs in presence of $\beq{k}$-cubic Dresselhaus SO couplings.

\section*{Acknowledgement}

D. Iizasa, M. Kohda, and J. Nitta are supported via Grant-in-Aid for Scientific Research (Grant No. 15H02099, No. 15H05854, No. 25220604, and No. 15H05699) by the Japan Society for the Promotion of Science.
D. Iizasa thanks the Graduate Program of Spintronics at Tohoku University, Japan, for financial support.
M. Kammermeier and U. Z\"ulicke acknowledge the support by the Marsden Fund Council from New Zealand government funding (contract no.\ VUW1713), managed by the Royal Society Te Ap\={a}rangi.

\appendix

\section{Derivation of the spin-orbit fields for general growth directions}\label{app:SOF}

In this appendix, we summarize the derivation of expressions (\ref{firstangular}) and (\ref{thirdangular}) for the spin-orbit fields of an arbitrarily oriented 2DEG as carried out in Ref.~\onlinecite{Kammermeier2016}.
\parag
Let us consider a 2D electron system that is confined by a quantum well grown along an arbitrary normal unit vector $\hat{\beq{ n}}=(n_x,n_y,n_z)$. 
In this system, the underlying basis vectors $\hat{\beq{x}}$, $\hat{\beq{y}}$, and $\hat{\beq{z}}$ correspond to the high-symmetry crystal directions [100], [010], and [001], respectively.
Pursuant to the 2D confinement, the electron wave vector $\beq{k}=(k_x,k_y,k_z)$ is decomposed as $\beq{k}=\beq{k}^\parallel+\beq{k}_n$, where $\beq{k}_n=\beq{\hat{n}}\,(\beq{\hat{n}}\cdot\beq{k})$ is pointing into the growth direction while $\beq{k}^\parallel$ lies in the 2DEG plane and, thus, obeys the constraint $\beq{\hat{n}}\cdot\beq{k}^\parallel=0$.
The effects of SO coupling are described by the Hamiltonian $\mathcal{H}_{\rm SO}=
\frac{\hbar}{2}(\beq{\Omega}_{\rm R}+\beq{\Omega}_{\rm D})\cdot\boldsymbol{\sigma}$ with the vector of Pauli matrices $\boldsymbol{\sigma}$.
The generic Rashba (R) and Dresselhaus (D) SO fields $\beq{\Omega}_{\rm R}$ and $\beq{\Omega}_{\rm D}$ can be written as\cite{Dyakonov1986,Kammermeier2016,KammermeierPHD}
\begin{align}
\boldsymbol{\Omega}_\text{R}={}&\frac{2\alpha}{\hbar}\left(\beq{k}^\parallel\times\beq{\hat{n}}\right),\quad
\boldsymbol{\Omega}_\text{D}={}\frac{2\gamma_{\rm D}}{\hbar} \,\boldsymbol{\nu},\label{eq:SOF}
\end{align}
where   
\begin{align}
\nu_x={}&\langle k_n^2\rangle\left[2n_x(n_yk^\parallel_y-n_zk^\parallel_z)+k^\parallel_x(n_y^2-n_z^2)\right]\notag\\
&+k^\parallel_x\left({k^\parallel_y}^2-{k^\parallel_z}^2\right)
\label{eq:dre_comp}
\end{align}
and similarly for $\nu_y$ and $\nu_z$ as obtained by cyclic index permutation.
The Rashba parameter $\alpha=\gamma_{\rm R}\langle\mathcal{E}_n\rangle$ 
depends on the electric field $\beq{\mathcal{E}}=\mathcal{E}_n\hat{\beq{n}}$ and the coefficients $\gamma_{\rm R}$ and $\gamma_{\rm D}$ constitute material specific parameters.
The expectation values 
denote the average over the ground-state wave function of the quantum well.
\parag
As demonstrated in Ref.~\onlinecite{Kammermeier2016}, a 2DEG can host persistent spin textures if and only if at least two growth-direction Miller indices agree in modulus.
Thus, without loss of generality we focus in the following on the growth direction given by the vector $\beq{\hat{n}}=(\eta,\eta,n_z)$, where $n_z=\sqrt{1-2\eta^2}$ and $\eta\in [1,1/\sqrt{2}]$.
This vector lies in the first quadrant of the [110]-[001] crystal plane as illustrated in Fig.~\ref{one}(a).
The relation to the polar angle $\theta\in [0,\pi/2]$, measured from the [001] axis, is given by $\eta=\sin(\theta)/\sqrt{2}$.
\parag
For the calculations in this paper it is more practical to choose a reference frame whose basis is adapted to the geometry of the 2DEG, i.e., one basis vector is aligned with the quantum-well growth direction while the others lie in the quantum-well plane.
Selecting $\hat{\beq{x}}=(n_z,n_z,-2\eta)/\sqrt{2}$, $\hat{\beq{y}}=(-1,1,0)/\sqrt{2}$, and $\hat{\beq{z}}\equiv\hat{\bf n}=(\eta,\eta,n_z)$ as in the main text [cf. Fig.~\ref{one}(a)], the transformation of $\mathcal{H}_{\rm SO}$ into the new basis is achieved by replacing $\beq{k}\mapsto  \mathcal{R}\beq{k}$ and $\beq{\sigma}\mapsto \mathcal{R}\beq{\sigma}$ using the rotation matrix
\begin{align}
\mathcal{R}={}&\frac{1}{\sqrt{2}}
\begin{pmatrix}
n_z&-1 &\sqrt{2}\eta \\
n_z&1 &\sqrt{2}\eta \\
-2\eta &0 &\sqrt{2}n_z \\
\end{pmatrix}.
\label{eq:rotmatrix}
\end{align}
After performing this transformation, the in-plane wave vector is given by only two components $\beq{k}^\parallel=(k^\parallel_x,k^\parallel_y)$ and we identify $\langle k_n^2\rangle\equiv\langle k_z^2\rangle$ and $\langle\mathcal{E}_n\rangle\equiv\langle\mathcal{E}_z\rangle$.
Henceforth, we suppress the superscript $\parallel$ and introduce polar coordinates for the in-plane wave vector, i.e., $\beq{k}=(k\cos\varphi,k\sin\varphi)$.
In a final step, the transformed total SO field is sorted with respect to their angular harmonic order in $\beq{k}$, where  $\beq{\Omega}_1$ and $\beq{\Omega}_3$ contain the first and third angular harmonics yielding the expressions (\ref{firstangular}) and (\ref{thirdangular}), respectively.

\section{Spin Diffusion Operator}\label{app1}

For a general growth direction with two Miller indices equal in modulus and arbitrary SO strengths, the spin diffusion operator $\sdo(\beq{q})$ in Eq.~(\ref{diffeq}) reads 
\begin{equation}
\sdo = 
\begin{pmatrix}
\tilde{\Lambda}_{xx}   &  \tilde{\Lambda}_{xy}  & \tilde{\Lambda}_{xz}\\
\tilde{\Lambda}_{xy}^\ast   &  \tilde{\Lambda}_{yy}  & \tilde{\Lambda}_{yz}\\
\tilde{\Lambda}_{xz}^\ast  &  \tilde{\Lambda}_{yz}^\ast  & \tilde{\Lambda}_{zz}\\
\end{pmatrix}
,
\end{equation}
with 
\begin{align}
\tilde{\Lambda}_{xx}=&\cfrac{4\pi^2}{\tau_0}\Bigg[\cfrac{1 - 18 \eta^2 + 105 \eta^4 - 144 \eta^6}{4}- \cfrac{ (1-9 \eta^2)n_z}{2} \Gamma_1\nonumber\\
&+\cfrac{\Gamma_1^2}{4} + \cfrac{1 + 10 \eta^2 - 15 \eta^4}{4} \Gamma_3^2 +\cfrac{\beq{q}^2}{\Qo^2}\Bigg],\\
\tilde{\Lambda}_{xy}=&\cfrac{4\pi^2}{\tau_0}\,i \sqrt{2}  (\eta - 3 \eta^3) \cfrac{q_y}{\Qo},\\
\tilde{\Lambda}_{xz}=&\cfrac{4\pi^2}{\tau_0}\Bigg\{\cfrac{\sqrt{2}(\eta-9\eta^5)n_z}{4}+
\left[  \cfrac{\sqrt{2}(\eta-3\eta^3)}{4} -i\cfrac{q_x}{\Qo} \right]\Gamma_1\nonumber\\
&-\cfrac{3(\eta-4\eta^3+3\eta^5)n_z}{2\sqrt{2}}\Gamma_3^2-i(1-9\eta^2)n_z\cfrac{q_x}{\Qo}\Bigg\},\\
\tilde{\Lambda}_{yy}=&\cfrac{4\pi^2}{\tau_0}\,\Bigg[\cfrac{1 + 6 \eta^2 - 15 \eta^4}{4}+\cfrac{(1+3\eta^2)n_z}{2}\Gamma_1+\cfrac{\Gamma_1^2}{4}\nonumber\\
&+\cfrac{1 + 10 \eta^2 - 15 \eta^4}{4}\Gamma_3^2+ \cfrac{\beq{q}^2}{\Qo^2}\Bigg],\\
\tilde{\Lambda}_{yz}=&-\cfrac{4\pi^2}{\tau_0}\,i(\Gamma_1 + n_z + 3 \eta^2 n_z)  \cfrac{q_y}{\Qo},\\
\tilde{\Lambda}_{zz}=&\cfrac{4\pi^2}{\tau_0}\Bigg[\cfrac{1 - 8 \eta^2 + 57 \eta^4 - 90 \eta^6}{2}+6 \eta^2 n_z\Gamma_1 +\cfrac{\Gamma_1^2}{2}\nonumber\\
&+\cfrac{1 - 8 \eta^2 + 21 \eta^4 - 18 \eta^6}{2}\Gamma_3^2+ \cfrac{\beq{q}^2}{\Qo^2}\Bigg],
\end{align}
where we used the basis vectors $\{\hat{\beq{x}},\hat{\beq{y}},\hat{\beq{z}}\}$ as defined in Sec.~\ref{sec:SOfields}.

\bibliographystyle{apsrev4-1}
\bibliography{bibspace}

\end{document}